\documentstyle[epsfig]{europhys}

\newif\ifboo \boofalse

\newcommand{\LI}{$^6$Li}
\newcommand{\bfr}{{\bf r}}

\newcommand{\bfR}{{\bf R}}
\newcommand{\bfE}{{\bf E}}
\newcommand{\bfx}{{\bf x}}
\newcommand{\bfy}{{\bf y}}

\newcommand{\bfk}{{\bf k}}

\newcommand{\bfp}{{\bf p}}

\newcommand{\bge}{\begin{equation}}
\newcommand{\ee}{\end{equation}}
\newcommand{\bgea}{\begin{eqnarray}}
\newcommand{\eea}{\end{eqnarray}}
\newcommand{\erw}[1]{\langle #1 \rangle}
\newcommand{\kint}{\int \frac{d^3\bfk}{(2\pi)^3}\:}

\begin{document}

\euro{}{}{-}{}
\Date{}
\shorttitle{F. Weig  }

\title{Optical detection of a BCS transition of Lithium--6 in harmonic traps}
\author{F.\ Weig and W. Zwerger}
\institute{Sektion Physik, 
Universit\"at M\"unchen, Theresienstr.\ 37/III,
D-80333 M\"unchen,\\ Germany}

\rec{}{}

\pacs{
\Pacs{03}{75.Fi}{Phase coherent atomic ensembles; quantum condensation
phenomena} 
\Pacs{32}{80Pj}{Optical cooling of atoms; trapping} 
\Pacs{05}{30Fk}{Fermion systems and electron gas} 
}

\maketitle

\begin{abstract}
We study the detection of a BCS transition within a sample of
Lithium--6 atoms confined in a harmonic trap. Using the
local density approximation we calculate the pair correlation
function in the normal and superfluid state at zero temperature. We
show that the softening of the Fermi hole associated with a BCS
transition leads to an observable increase in the intensity
of off--resonant light scattered from the atomic cloud at small angles.
\end{abstract}

The experimental realization of Bose-Einstein-Condensation in dilute
Alkali vapors~\cite{bib:bec} stimulated a variety of experimental and
theoretical work in the field of quantum gases.  Recently, attention
was also drawn to the special properties of ultracold fermionic
gases. In particular the absence of a $s$--wave scattering interaction
in a spin polarized system would allow to realize an almost perfect
example of a noninteracting Fermi gas, of which first indications were
observed in an experiment with $^{40}$K atoms~\cite{bib:jin}. In the
case of an attractive interaction between different hyperfine levels
as occurs in {\LI}~\cite{bib:aket}, the ground state will be a BCS
superfluid. Stoof {\it et al.}~\cite{bib:stoof1} first proposed to
look for this phase transition in a mixture of two hyperfine states of
\LI, where due to the extraordinary large $s$--wave scattering length
the transition temperature turns out to be in an observable range of
about $100$nK. Their work was further expanded by Modawi and
Leggett~\cite{bib:modawi}, including all three hyperfine states that
can simultaneously be enclosed in a magnetic trap.

An experimental realization of the BCS--transition in magnetic traps
is aggravated by particular conditions regarding the life time of the
sample, equal occupation of the hyperfine states and a large magnetic
offset field to obtain the maximal scattering length of $a = -1140$\AA
\ for lithium~\cite{bib:stoof3}, so that cooling of the sample in a
pure optical trap might be advantageous. Besides these difficulties,
the most favorable way to detect a BCS--transition in future
experiments is still a matter of discussion. In fact, the differences
in the density and momentum distribution between the normal and
superfluid phases are tiny~\cite{bib:stoof2}, so that unlike the case
of bosonic gases, time--of--flight measurements will not be able to
indicate the phase transition. As possible observable signatures
collective excitations~\cite{bib:baranov4,bib:bruun2}, changes in the
line width and line shift in light scattering
experiments~\cite{bib:ruostekoski} and anomalous moments of
inertia~\cite{bib:farine} have been suggested. In this paper we show
that the BCS transition in a cloud of cold atoms may be detected
directly by off--resonant light scattering. This suggestion was made
independently by Zhang {\it et al.}~\cite{bib:zhang}, however our
results for the signature of the transition are quite different from
theirs, see our discussion at the end of the paper. Recent
measurements of the dynamic structure factor in Bose
condensates~\cite{bib:ketterle} show that even the time
dependent pair
correlation functions have indeed become accessible by experiments.

We consider a system of fermionic atoms in two equally occupied
hyperfine states trapped in an isotropic harmonic potential $V(\bfx) =
\frac{1}{2}m\omega^2x^2$. Atoms in different hyperfine states interact
via an attractive contact interaction $V(\bfx,\bfy) = V_0
\delta(\bfx-\bfy)$. The corresponding Hamiltonian is 

\begin{equation}
\hat{H} = \sum_\sigma \int d^3x \: \Psi^\dagger_\sigma(\bfx)
{\mathcal{H}}_0 \Psi_\sigma(\bfx) + \frac{V_0}{2} \sum_\sigma \int
d^3x  \Psi^\dagger_\sigma(\bfx) \Psi^\dagger_{-\sigma}(\bfx)
\Psi_{-\sigma}(\bfx) \Psi_\sigma(\bfx).
\end{equation}

where $\Psi_\sigma(\bfx)$ is a fermionic field operator which destroys
a particle in hyperfine state $\sigma$ at location $\bfx$ and
${\mathcal H}_0= -\hbar^2/2m \nabla^2 + V(\bfx)-\mu$ denotes the
single particle Hamiltonian. The diagonalization of this Hamilton
operator in mean field approximation is equivalent to solving
the Bogoliubov--de Gennes equations, originally formulated for
inhomogeneous superconductors~\cite{bib:degennes}

\begin{eqnarray}
\label{eq:BdGG}
E_\bfk u_\bfk(\bfx) &=& \; \; \: [{\mathcal H}_0 + W(\bfx)]u_\bfk(\bfx)
+ \Delta(\bfx) v_\bfk(\bfx) \\
\label{eq:BdGG2}
E_\bfk v_\bfk(\bfx) &=& - [{\mathcal H}_0 + W(\bfx)]v_\bfk(\bfx) +
\Delta(\bfx) u_\bfk(\bfx).
\end{eqnarray} 

Here we introduced the Hartree potential $W(\bfx) = V_0
\erw{\Psi^\dagger_\sigma(\bfx) \Psi_\sigma(\bfx)}$ and the pair
potential $\Delta(\bfx) = -V_0 \erw{\Psi_\sigma(\bfx)
\Psi_{-\sigma}(\bfx)}$, which both depend on the position in the
trap. As in the conventional BCS theory the function $v_\bfk(\bfx)$
can be interpreted as the amplitude for the occupation of a Cooper
pair with excitation energy $E_\bfk$. The amplitudes are normalized by

\begin{equation}
\label{eq:BdGnorm}
\sum_\bfk \left( u^{\phantom{*}}_\bfk(\bfx) u^*_\bfk(\bfy)
+ v^{\phantom{*}}_\bfk(\bfx) v^*_\bfk(\bfy) \right) = \delta(\bfx - \bfy). 
\end{equation}

We will not attempt to provide a complete solution of these equations
which is quite involved numerically and also requires a nontrivial
regularization procedure for the pseudo
potential~\cite{bib:bcdb}. Instead, we use the local density
approximation (LDA), which relies on the fact that both the Fermi
wavelength $\lambda_F$ and the BCS coherence length $\xi_0$ are much
smaller than the typical scale $x_0 = \sqrt{\hbar/m\omega}$, on which
the harmonic external potential varies. The results of the LDA can be
obtained from the Bogoliubov--de Gennes equations by using an ansatz
$u_\bfk(\bfx) = \exp(i\bfk \bfx) \tilde{u}_\bfk(\bfx)$, $v_\bfk(\bfx)
= \exp(i\bfk \bfx) \tilde{v}_\bfk(\bfx)$ with slowly varying
amplitudes $\tilde{u}_\bfk(\bfx)$, $\tilde{v}_\bfk(\bfx)$.  Inserting
this into the Bogoliubov--de Gennes equations and neglecting all
derivatives of $\tilde{u}_\bfk(\bfx)$ and
$\tilde{v}_\bfk(\bfx)$\footnote{The LDA can also be interpreted as the
semi-classical $\hbar \rightarrow 0$ limit of an Ansatz $u_\bfk(\bfx)
= e^{i\bfp \bfx/\hbar} \tilde{u}_\bfk(\bfx)$.}, the excitation energy
is the familiar
$E_\bfk(\bfx)=\sqrt{(\varepsilon_\bfk-\mu(\bfx))^2+\Delta^2(\bfx)}$
with the local chemical potential $\mu(\bfx)=\mu - V(\bfx) -
W(\bfx)$. Similarly the amplitudes take the standard BCS form

\begin{equation}
\label{eq:ampsf}
\tilde{u}_\bfk(x) = \sqrt{\frac{1}{2} \left(1+
\frac{\xi_\bfk(x)}{E_\bfk(x)} \right)},  \; \; \; \; \; \; 
\tilde{v}_\bfk(x) = \sqrt{\frac{1}{2} \left(1-
\frac{\xi_\bfk(x)}{E_\bfk(x)} \right)}. 
\end{equation}

with the reduced single particle energies $\xi_\bfk(\bfx) =
\varepsilon_\bfk - \mu(\bfx)$.  In order to determine the local
values of $\mu(\bfx)$ and $\Delta(\bfx)$ we numerically solve the
standard gap equation together with that for the total number of
particles. The ultraviolet divergence in the gap
equation which arises from the assumption of a contact potential with
no extrinsic cutoff in energy of the attractive interaction, can be
eliminated by using the standard relation between the bare interaction
parameter $V_0$ and the low energy effective interaction $g =
4\pi\hbar^2 a / m$, which is determined by the scattering length
$a$. The resulting density distribution and
local pair potential are in very good agreement with those found 
earlier~\cite{bib:stoof2,bib:farine,bib:bcdb}, with a typical cloud 
radius $R_{T}\approx v_{F}/\omega$. For particle numbers of order
$6\cdot 10^{5}$ used in our calculations, the LDA is in fact an
excellent approximation as was shown by Bruun {\it et al.}~\cite{bib:bcdb}. 

We now turn to the calculation of the pair correlation function. Using
the Wick--theorem the pair correlation function $g(\bfx,\bfy)$
separates into a normal part $g^N(\bfx,\bfy)$ and an anomalous part
$g^A(\bfx,\bfy)$:

\begin{equation}
g({\bfx},\bfy) = \erw{\hat{n}({\bfx})\hat{n}(\bfy)} -
\erw{\hat{n}({\bfx})}\erw{\hat{n}(\bfy)} = g^N({\bfx},\bfy) +
g^A({\bfx},\bfy).
\end{equation} 

The normal part correlation function consists of the autocorrelation
part and the anti--correlation between fermions in the same spin
states. The anomalous correlation function is nonzero only in the
superfluid phase, when anomalous expectation values occur due to the
formation of pairs. At zero temperature and within the LDA the normal
and anomalous parts of the correlation function are given by

\bgea
\label{eq:corrn}
g^N({\bfx},\bfy) &=& n(\bfx) \delta(\bfx - \bfy) - 2\, \left( \kint
e^{i\bfk(\bfx-\bfy)} \tilde{v}_\bfk(\bfx)\,
\tilde{v}_\bfk(\bfy)\right)^2. \\
\label{eq:corra}
g^A({\bfx},\bfy) &=& 2 \,F(\bfx,\bfy) \, F^*(\bfy,\bfx).
\eea

where $F(\bfx,\bfy)$ is the anomalous pair amplitude 

\begin{equation} 
\label{eq:corraG}
F(\bfx,\bfy) = \kint e^{i\bfk(\bfx-\bfy)} \tilde{u}_\bfk(\bfx)
\,\tilde{v}_\bfk(\bfy).
\end{equation}

The pair amplitude diverges for $r = |\bfx-\bfy| \rightarrow 0$ like
$m\Delta(y)/4\pi \hbar^2 r$, which is an artefact of the zero range 
interaction. Unlike the pair potential this divergence
cannot be compensated by a divergence in the coupling
constant. A rough estimate for the local pair correlator is obtained
by taking $F(\bfx,\bfx) = \Delta(\bfx)/g$ with $g$ as the 
renormalized interaction constant, leading to

\begin{equation}
\label{eq:SoftHole}
g(\bfx,\bfx) = -\frac{n^2(\bfx)}{2} + 2 \frac{\Delta^2(\bfx)}{g^2}.
\end{equation}

While the scattered intensity calculated below is rather insensitive
to the precise value of $g(\bfx,\bfx)$, this relation shows explicitly 
that the positive anomalous correlations
associated with a BCS--transition reduce the effect of the Fermi hole
in the pair correlation function of fermions.

In the normal state,
where $\Delta(\bfx)= 0$, the anomalous correlations vanish and we
obtain the following inhomogeneous generalization of a well--known
result for Fermi gases:

\begin{equation}
\label{eq:corrnf}
g^N(\bfx,\bfy) = n(z)\,\delta(\bfr)-\frac{1}{2}\, \left[ \frac{3\,
n(z)}{(k_F(z)r)^3} \Big( \sin(k_F(z)r)-k_F(z) r \cos(k_F(z) r)\Big) 
\right]^2.
\end{equation}
 
\begin{figure} 
{\par\centering {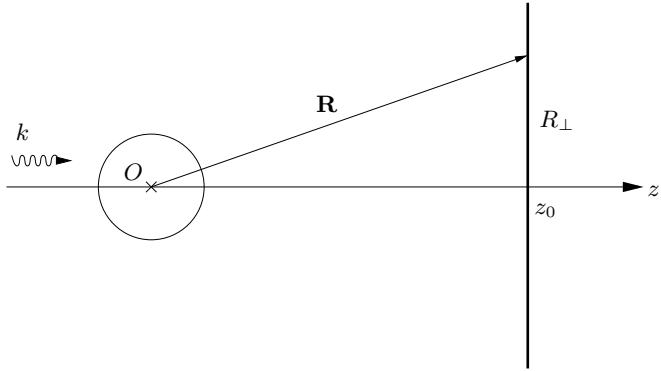}\par}
\caption[Scattering geometry]{The scattering geometry used in this
paper: A monochromatic light beam with wave vector $\bfk$ scatters at
the atomic trap placed at the origin. In a minimal distance $z_0$ from
the trap an observation screen is placed. The scattered intensity is
measured at distance $R_\bot$ from the center of the observation
screen.
\label{fig:scattergeo}}
\end{figure}
 
Here we introduced $z =$ Max$(x,y)$, $r=|\bfx-\bfy|$ and the local
Fermi wave vector $k_F(z)= \sqrt{2m\mu(z)/\hbar^2}$. The
correlation function shows the typical Friedel oscillations with
period $\lambda_F/2$ and an algebraic decay $r^{-4}$ caused by the
Fourier transform of the step in the Fermi distribution at zero
temperature. In the superfluid regime with $\Delta(\bfx)\not= 0$, the pair
correlations can only be calculated numerically. Reflecting the tiny
differences of the density distributions, the normal correlation
functions differ only at large distances, where their absolute values
are negligible. The anomalous correlation function also exhibits Friedel
oscillations and an exponential decay on a scale $k_F^{-1}$. Naively
one would expect a decay with the BCS coherence length
 $\xi_0=\hbar v_{F}/\Delta_{0}$ as in the homogeneous
case, but the inhomogenity of the amplitudes $\tilde{u}_k(\bfx)$ and
$\tilde{v}_k(\bfx)$ blocks this effect. Overall, the magnitude of the anomalous
correlation function is about $10\%$  of the 
normal correlations although it is formally a very small
effect of order $(\Delta_0/\varepsilon_F)^2\approx 0.01$  
for realistic parameters in~\LI.

In the following,  we will see
that due to an almost perfect cancellation of the autocorrelation and
the normal correlation contributions in an off--resonance light
scattering experiment, the effect of the anomalous correlations is
quite appreciable. The scattering geometry is shown in
Figure~\ref{fig:scattergeo}. The incoming laser beam with amplitude
$\bfE_L$ and wave vector $k$ produces an atomic polarization ${\bf
P}(\bfr)$ which for a large detuning $\delta$ of the laser frequency
from the resonance is given by

\begin{equation}
{\bf P}(\bfr) = - \frac{d^2 {\bf E}(\bfr)}{\hbar \delta} n(\bfr).
\end{equation}

Here $d$ is the matrix element of the atomic dipole moment and
$n(\bfr)$ the gas density in the trap. We calculate the scattered
field ${\bf E}_{sc}$ in first order Born approximation. Assuming
that the incoming laser field is removed by the dark ground technique
and neglecting the temporal variation due to the excitation
processes, the measured intensity at position
$R_\bot$ on the screen is given by~\cite{bib:java1,bib:demarco}

\begin{equation} 
I(\bfR) = \erw{\bfE_{sc}(\bfR) \bfE_{sc}(\bfR)} = \frac{9 I_L}{16(k R^2
\delta)} \left(F_n(R_\bot) + F_c(R_\bot)\right),
\end{equation}
  
where $I_L$ is the laser intensity. The first contribution to the
intensity is essentially the Fourier transform of the density
distribution in the trap

\begin{equation}
\label{eq:IntRho}
F_n(R_\bot) = \left| \int d^3r \: e^{i k ({\bf \hat{R}}-{\bf
\hat{e}}_z)\cdot \bfr} n(\bfr) \right|^2.
\end{equation}

where ${\bf \hat{R}}$ and ${\bf\hat{e}}_z$ are unit vectors in the
direction of $\bfR$ and in $z$--direction respectively. The second
contribution to the intensity contains the pair correlation function:

\begin{equation}
\label{eq:IntCorr}
F_c(R_\bot) = \int\int d^3r d^3r' \: e^{i k ({\bf \hat{R}}-{\bf
\hat{e}}_z)\cdot (\bfr-{\bf r'})} g(\bfr,{\bf r'}).
\end{equation}  

The contribution $F_n(R_\bot)$ can be described as simple coherent
scattering from a circular hole with an amplitude proportional to
$N^2$ and indeed our numerical results show the typical Bessel function
behaviour~\cite{bib:bornwolf}. From the first minimum in
the scattered intensity, which occurs at $1.22 R \lambda /d$, we can
extract an effective radius $d/2$ of the circular opening. Due to the
variation of the density through the atomic cloud this radius turns out
to be about half the Thomas--Fermi radius $R_{T}$ of the trap. For the
numerical calculations we use a trap frequency $\omega = 2\pi\times
144$Hz and chemical potential $\mu_0 = 110.5 \hbar\omega$ as typical
parameters~\cite{bib:stoof2}. This corresponds to an oscillator length
of $x_0 \approx 3.4\mu$m, a Thomas--Fermi radius of about $15x_0$ and
a Fermi wavelength $k_F^{-1} \approx 0.2 x_0$ in the center of the
trap. If we place the screen in a distance of $z_0 = 2$ cm from the
trap, the first minimum in the scattered intensity is only $0.3$ mm
away from the screen center.

For the intensity contribution due to the correlations within the gas,
the typical length scale is not the radius $R_T$ of the trap, but the
much smaller Fermi wavelength $\lambda_F$. Since the scattering angle
is proportional to the inverse typical length scale, the scattered
intensity due to the correlations forms a much wider light cone. The
Fermi wavelength is the typical scale for all parts of the correlation
function. The normal part of the pair correlation function consists of
two components, namely the autocorrelation function and the Friedel
oscillating part which carry opposite signs. The positive
autocorrelation leads to a quasi constant intensity proportional to
$N/R^2$, whereas the contribution of the Fermi hole with the Friedel
oscillations is negative. The result is a strong suppression of the
scattered intensity at small angles, reflecting the Fermi hole of the
pair correlation function. While this effect could be used to detect
the Fermi degeneracy in normal systems~\cite{bib:demarco}, the small
positive contribution of the anomalous correlations also becomes
significant.

\begin{figure}
{\par\centering {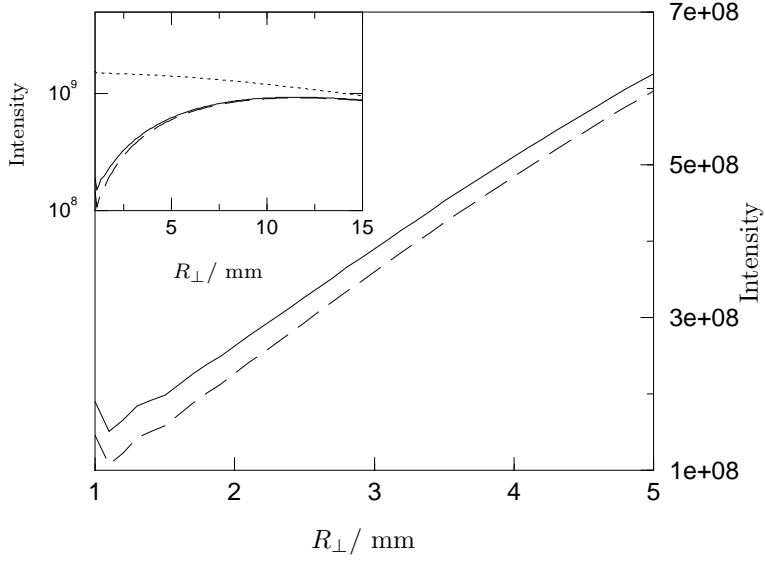}\par}
\caption[Scattered Intensity]{The total scattered intensity
distribution in arbitrary units with the distance $R_\bot$ in mm from the
center of the observation screen. The dashed line denotes the normal
fluid state, the solid line the superfluid state. The inset also shows
the intensity scattered from an ideal classical gas.  
\label{fig:xgscatter}}
\end{figure}

To demonstrate the effect of the BCS correlations, we numerically 
evaluated the total intensity
on a screen with distance $z_0 = 2$cm from the center of the atomic
cloud. The anomalous pair correlation function was determined by a
refined Simpson routine, the integrals in (\ref{eq:IntCorr}), only one
of which can be performed analytically, by a Monte Carlo method. We
evaluated the pair correlation function at $200$ million coordinate
pairs $(\bfr,{\bf r'})$ within the trap. The result is shown in
figure~\ref{fig:xgscatter}, where the scattered intensity from a
normal fluid and a superfluid gas of $^6$Li atoms are compared. For
small distances $R_\bot$ from the center of the screen the superfluid
phase transition significantly rises the intensity. The inset shows
the `Fermi hole', the drastic reduction of the observed intensity of
the fermionic gas in comparison with the intensity scattered from an
ideal gas of atoms, where just the autocorrelation function
contributes. The typical error bars on the curves as determined by the
Monte Carlo method were below $5\%$ for the normal part and below
$10\%$ for the anomalous part.

\begin{figure}
{\par\centering {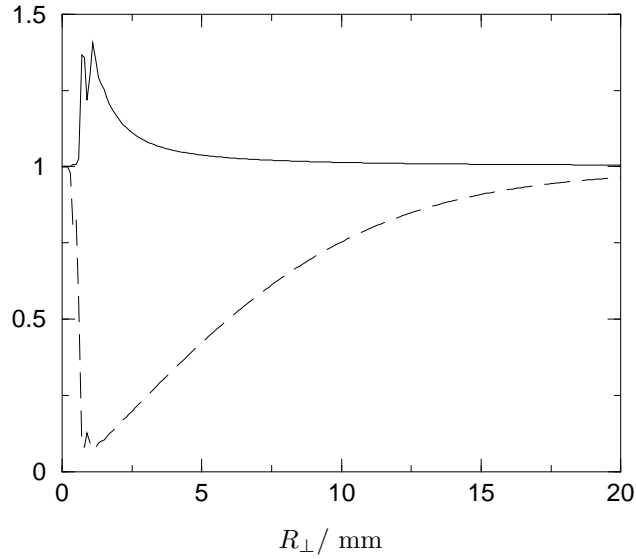}\par}
\caption[Proportional Intensity]{The ratio of the measured intensities
for different states with the distance $R_\bot$ in mm from the center
of the screen. The dashed line denotes the ratio of normal fluid
fermionic gas to an ideal classical gas, the solid line the ratio of
superfluid fermionic gas to the normal fluid gas.
\label{fig:xscatterprop}}
\end{figure}

As is shown in figure~\ref{fig:xscatterprop} the measured
intensity of a normal fluid Fermi gas drops by almost $90\%$ in
comparison with the ideal gas at zero temperature. This value for the
normal fluid gas, however, is raised by almost $40\%$ by the superfluid
transition at small scattering angles, reflecting the softening of the
Fermi hole by the anomalous correlations as in
eq.(\ref{eq:SoftHole}). The peaks that occur within this graph result
from side maxima of the central coherent scattering peak, which for
our system is about seven orders of magnitude stronger than the
typical intensity caused by the correlations. Although the rise in the
intensity will be smaller at finite temperatures, it should still be
measurable. It is remarkable how the anomalous correlations
qualitatively change the behaviour of the scattered intensity: at a fixed
position one would first observe a drop in the intensity with falling
temperature caused by Fermi statistics and below $T_c$ a subsequent
rise. 

Very recently, the suggestion to detect a BCS transition via 
off-resonant light scattering has also been
made by Zhang {\it et al.}~\cite{bib:zhang}. They found that the
scattered light cone widens from an angle of order
$\lambda/R_{T}$ in the normal state to an angle of order 
$\lambda/\lambda_{F}$ in the presence of anomalous correlations.
However in this work the normal part $g^N(\bfx,\bfy)$ of the pair
correlation function was neglected and the authors assumed $R \approx
R_\bot$ to perform the integration over the $z$--coordinate in
equations (\ref{eq:IntRho}) and (\ref{eq:IntCorr}). Therefore the
cancellation between the autocorrelation and the normal part of
$g(\bfx,\bfy)$ which is responsible for the Fermi hole and its 
subsequent softening by the BCS correlations is not seen in their 
results.

In conclusion we have performed a quantitative study of the
possibility to detect a superfluid transition within a fermionic gas,
e.g. $^6$Li, by measuring the intensity of off-resonant light
scattered from the atomic trap. The intensity outside the central
coherent scattering cone is dominated by the contributions of the pair
correlation function of the system. The pair correlation function of a
superfluid gas contains an anomalous part, which rises the observed
intensity by as much as $40\%$ compared to the degenerate Fermi gas,
thus providing a clear signature for the onset of superfluid
correlations.


\begin{thebibliography}{99}
\bibitem{bib:bec} Anderson M.H. {\it et al}, Science {\bf 269}, 198
(1995), Davis K.B. {\it et al}, PRL {\bf 75}, 3969 (1995), 
Bradley C.C. {\it et al}, PRL {\bf 75}, 1687 (1995).
\bibitem{bib:jin} DeMarco B., Bohn J.L., Burke J.P. Jr., Holland M.,
Jin D.S., PRL {\bf 82}, 4208 (1999).
\bibitem{bib:aket} Abraham E.R.I., McAlexander W.I., Gerton J.M.,
Hulet R.G., C\^ot\'e R.,\\ Dalgarno A.,  PRA {\bf 55}, R3299 (1997).
\bibitem{bib:stoof1} Stoof H.T.C., Houbiers M., Sackett C.A., Hulet
R.G., PRL {\bf 76}, 10 (1996). 
\bibitem{bib:modawi} Modawi A.G.K., Leggett A.J., J. Low
Temp. Phys. {\bf 109}, 625 (1997).  
\bibitem{bib:stoof3} Houbiers M., Stoof H.T.C., McAlexander W.I.,
Hulet R.G., PRA {\bf 57}, R1497 (1998).
\bibitem{bib:stoof2} Houbiers M., Ferweda R., Stoof H.T.C.,
McAlexander W.I., Sackett C.A., Hulet R.G., PRA {\bf 56}, 4864 (1997).
\bibitem{bib:baranov4} Baranov M.A., Petrov D.S., cond-mat/9901108.
\bibitem{bib:bruun2} Bruun G.M., Clark C.W., cond-mat/9906392.
\bibitem{bib:ruostekoski} Ruostekoski J., cond-mat/9902324.
\bibitem{bib:farine} Farine M., Schuck P., Vinas X., cond-mat/9901241. 
\bibitem{bib:zhang} Zhang W., Sackett C.A., Hulet R.G., PRA {\bf 60},
504 (1999).
\bibitem{bib:ketterle} Stenger J., Inouye S., Chikkatur A.P.,
Stamper-Kurn D.M., Pritchard D.E., Ketterle W., PRL {\bf 82}, 4569 (1999).  
\bibitem{bib:degennes} de Gennes, {\it Superconductivity of Metals and
Alloys}, Addison--Wesley, 1966. 
\bibitem{bib:bcdb} Bruun G., Castin Y., Dum R., Burnett K.,
cond-mat/9810013.
\bibitem{bib:java1} Javanainen J., Ruostekoski J., PRA {\bf 52}, 3033
(1995).
\bibitem{bib:demarco} DeMarco B., Jin D.S., PRA {\bf 58}, R4267
(1998), compare also Busch T., Anglin J.R., Cirac J.I., Zoller P., EPL
{\bf 44}, 1 (1998).
\bibitem{bib:bornwolf} Born M., Wolf E., {\it Principles of Optics }
 6ed., Pergamon Press, 1980.

\end{thebibliography}
\end{document}